\documentclass[information,article,accept,moreauthors,pdftex,10pt,a4paper]{Definitions/mdpi}

\usepackage{amsmath,amssymb,amsfonts}
\usepackage{algorithmicx}
\usepackage[algo2e]{algorithm2e}

\usepackage{pseudocode}

\usepackage{mathtools}
\usepackage{algorithm}
\usepackage{rotating}
\usepackage{breakurl}
\usepackage{array}
\usepackage{graphicx}
\usepackage{color}
\usepackage{textcomp}
\usepackage{esvect}
\usepackage{amsmath,stackengine}

\setitemize{parsep=6pt,itemsep=0pt,leftmargin=*,labelsep=5.5mm}
\setenumerate{parsep=6pt,itemsep=0pt,leftmargin=*,labelsep=5.5mm}
\setlist[description]{itemsep=0mm}
\setitemize{parsep=6pt,itemsep=0pt,leftmargin=*,labelsep=5.5mm,align=parleft}
\setenumerate{parsep=6pt,itemsep=0pt,leftmargin=*,labelsep=5.5mm,align=parleft}

\usepackage{soul}
\usepackage{booktabs}
\usepackage{multirow}
\usepackage{microtype}
\makeatletter

\titleformat{\subparagraph}
{\normalfont\normalsize\itshape}{\thesubparagraph}{1em}{}
\titlespacing*{\subparagraph}
{0pt}{1.25ex plus 1ex minus .2ex}{1ex plus .2ex}

\newcommand{\RNum}[1]{%
\textup{\uppercase\expandafter{\romannumeral#1}}%
}

\newcommand{\qu}[1]{``#1''}

\DeclareMathOperator*{\argmax}{arg\,max}

\firstpage{1} 
\pubvolume{11}
\issuenum{6}
\articlenumber{318}
\pubyear{2020}
\copyrightyear{2020}
\history{Received: 6 May 2020; Accepted: 10 June 2020; Published: 12 June 2020}
\updates{yes} 

\Title{HMIC: Hierarchical Medical Image Classification, A~Deep Learning Approach}

\Author{Kamran Kowsari$^{1,2,3,}$*\href{https://orcid.org/0000-0002-6451-4786}{\orcidicon}, Rasoul Sali~$^{1}$\href{https://orcid.org/0000-0003-3995-0735}{\orcidicon}, Lubaina Ehsan $^{4}$\href{https://orcid.org/0000-0001-6307-2106}{\orcidicon}, William Adorno$^{1}$, Asad Ali $^{5}$, Sean Moore $^{4}$\href{https://orcid.org/0000-0003-1150-6632}{\orcidicon}, Beatrice Amadi $^{6}$, Paul Kelly $^{6,7}$\href{https://orcid.org/0000-0003-0844-6448}{\orcidicon}, Sana Syed $^{4,5,8,}$*\href{https://orcid.org/0000-0003-0954-0583}{\orcidicon} and Donald~Brown~$^{1,8,}$*\href{https://orcid.org/0000-0002-9140-2632}{\orcidicon}}

\AuthorNames{Kamran Kowsari, Rasoul Sali, Lubaina Ehsan, William Adorno III, S. Asad Ali, Sean R. Moore, Beatrice C. Amadi, Paul Kelly, Sana Syed and Donald~Brown}

\address{%
$^{1}$ \quad Department of Systems and Information Engineering, University of Virginia, Charlottesville, VA 22904, USA; rs8wa@virginia.edu (R.S.); wa3mr@virginia.edu (W.A.)\\
$^{2}$ \quad Office of Health Informatics and Analytics, University of California, Los Angeles~(UCLA), CA 90095, USA\\
$^{3}$ \quad Sensing  Systems  for  Health  Lab, University of Virginia, Charlottesville, VA 22911, USA\\
$^{4}$ \quad Department of Pediatrics, School of Medicine,  University  of  Virginia,  Charlottesville,  VA {22903}, USA; 
 LE7JG@hscmail.mcc.virginia.edu (L.E.); sean.moore@virginia.edu (S.M.)\\
$^{5}$ \quad Department of Paediatrics and Child Health, The Aga Khan University, Karachi 74800, Pakistan; asad.ali@aku.edu\\
$^{6}$ \quad Tropical  Gastroenterology  and  Nutrition  Group,  University  of  Zambia School of Medicine, {32379 Lusak}, Zambia; beatriceamadi@ymail.com (B.A.); m.p.kelly@qmul.ac.uk (P.K.)\\
$^{7}$ \quad Blizard  Institute,  Barts  and  The  London  School  of  Medicine,  Queen Mary University of London, London~E1~4NS, UK\\ 
$^{8}$ \quad School of Data Science, University of Virginia, Charlottesville, VA 22904, USA\\}

\corres{Correspondence: kk7nc@virginia.edu (K.K.); sana.syed@virginia.edu (S.S.); deb@virginia.edu (D.B.); Tel.:~+1-202-812-3013 (K.K.)}

\abstract{Image classification is central to the big data revolution in medicine. Improved information processing methods for diagnosis and classification of digital medical images have shown to be successful via deep learning approaches. As this field is explored, there are limitations to the performance of traditional supervised classifiers. This paper outlines an approach that is different from the current medical image classification tasks that view the issue as multi-class classification. We performed a hierarchical classification using our Hierarchical Medical Image classification (HMIC) approach. HMIC uses stacks of deep learning models to give particular comprehension at each level of the clinical picture hierarchy. For testing our performance, we use biopsy of the small bowel images that contain three categories in the parent level~(Celiac Disease, Environmental Enteropathy, and histologically normal controls). For the child level, Celiac Disease Severity is classified into 4 classes~(I, IIIa, IIIb, and IIIC).}

\keyword{deep Learning, hierarchical classification, hierarchical medical image classification, medical imaging}

\begin{document}
\section{Introduction and Related Works}\label{sec:Introduction} 

Automatic diagnosis of diseases based on medical image categorization has become increasingly challenging over the last several years~\cite{sali2019celiacnet,kowsari2019diagnosis,kowsari2020Diagnosis}. Areas of research involving deep learning architectures for image analysis have grown in the past few years with an increasing interest in their exploration and understanding of the domain application~\cite{info10040150,litjens2017survey,nobles2018identification,zhai2016doubly,kowsari2020Diagnosis}. Deep learning models achieved state-of-the-art results in a wide variety of fundamental tasks such as image classification in the medical domain~\cite{hegde2019comparison,zhang2018patient2vec}. This growth has raised questions regarding classification of sub-types of disease across a range of disciplines including Cancer (e.g., stage of cancer), Celiac Disease (e.g., Marsh Score Severity Class), and Chronic Kidney Disease (e.g., Stage 1--5) among others~\cite{pavik2013secreted}. Therefore, it is important to not just label medical images-based specialized areas, but to also organize them within an overall field (i.e.,~name of disease) with the accompanying sub-field (i.e., sub-type of disease) which we have done in this paper via Hierarchical Medical Image Classification~(HMIC). Hierarchical models also combat the problem of unbalanced medical image datasets for training the model and have been successful for other domains~\cite{kowsari2017hdltex,dumais2000hierarchical}.

In the literature, few efforts have been  made  to  leverage  the  hierarchical  structure  of  categories. Nevertheless, hierarchical models have shown better performance compared to flat models in image classification across multiple domains~\cite{yan2019hierarchical, seo2019hierarchical, ranjan2018hierarchical}. These models exploit the hierarchical structure of object categories to decompose the classification tasks into multiple steps. Yan et al. proposed HD-CNN by embedding deep CNNs into a category hierarchy~\cite{yan2019hierarchical}. This model separates easy classes using a coarse category classifier while distinguishing difficult classes using  fine  category  classifiers. In~a~CNN, shallow layers capture low-level features while deeper layers capture high level ones. Zhu and Bain proposed Branch Convolutional Neural Network (B-CNN)~\cite{zhu2017b} based on this characteristic of CNNs. This model instead of employing different classifiers for different levels of class hierarchy, exploits the hierarchical structure of layers in a CNN and embeds different levels of class hierarchy on a single CNN. B-CNN outputs multiple predictions ordered from coarse to fine along concatenated convolutional layers corresponding to hierarchical structure of the target classes. Sali et al. employed B-CNN model for the classification of gastrointestinal disorders on histopathological images~\cite{sali2020hierarchical}.

Our paper uses the HMIC approach for assessment of small bowel enteropathies; Environmental Enteropathy (EE) versus Celiac Disease (CD) versus histologically normal controls. EE is a common cause of stunting in Low-to-Middle Income Countries (LMICs), for which there is no universally accepted, clear diagnostic algorithms or non-invasive
biomarkers for accurate diagnosis \cite{syed2016environmental}, making this a critical priority \cite{naylor2015environmental}. Linear growth failure (or stunting) is associated with irreversible
physical and cognitive deficits, with profound developmental implications \cite{syed2016environmental}. Interestingly, CD, a common cause of stunting in the United States, with an estimated $1$\% prevalence, is an autoimmune disorder caused by a gluten sensitivity \cite{husby2012european} and has many shared histological features with EE (such as increased inflammatory cells and villous blunting) \cite{syed2016environmental}. This resemblance has led to the major challenge of differentiating clinical biopsy images for these similar but distinct diseases. CD severity is further assessed via Modified Marsh Score Classification. It takes into account the architecture of the duodenum as having finger-like projections (called \qu{villi}) which are lined by cells called epithelial cells. Between the villi are crevices called crypts that contain regenerating epithelial cells. Normal villus to crypt ratio is between 3:1 and 5:1 and a healthy duodenum (first part of the small intestine) has no more than 30 lymphocytes interspersed per 100 epithelial cells within the villus surface layer (epithelium). Marsh I  comprises of normal villus architecture with an increase in the number of intraepithelial lymphocytes. Marsh II has increased intraepithelial lymphocytes along with crypt hypertrophy (crypts appear enlarged). This is usually rare since patients typically rapidly progress from Marsh I to IIIa. Marsh III is sub-divided into IIIa (partial villus atrophy), Marsh IIIb (subtotal villus atrophy) and Marsh IIIc (total villus atrophy) along with crypt hypertrophy and increased intra-epithelial lymphocytes. Finally, in Marsh IV, villi are completely atrophied \cite{fasano2001current}. 

\begin{figure}[H]
    \centering
    \includegraphics[width=0.85\columnwidth]{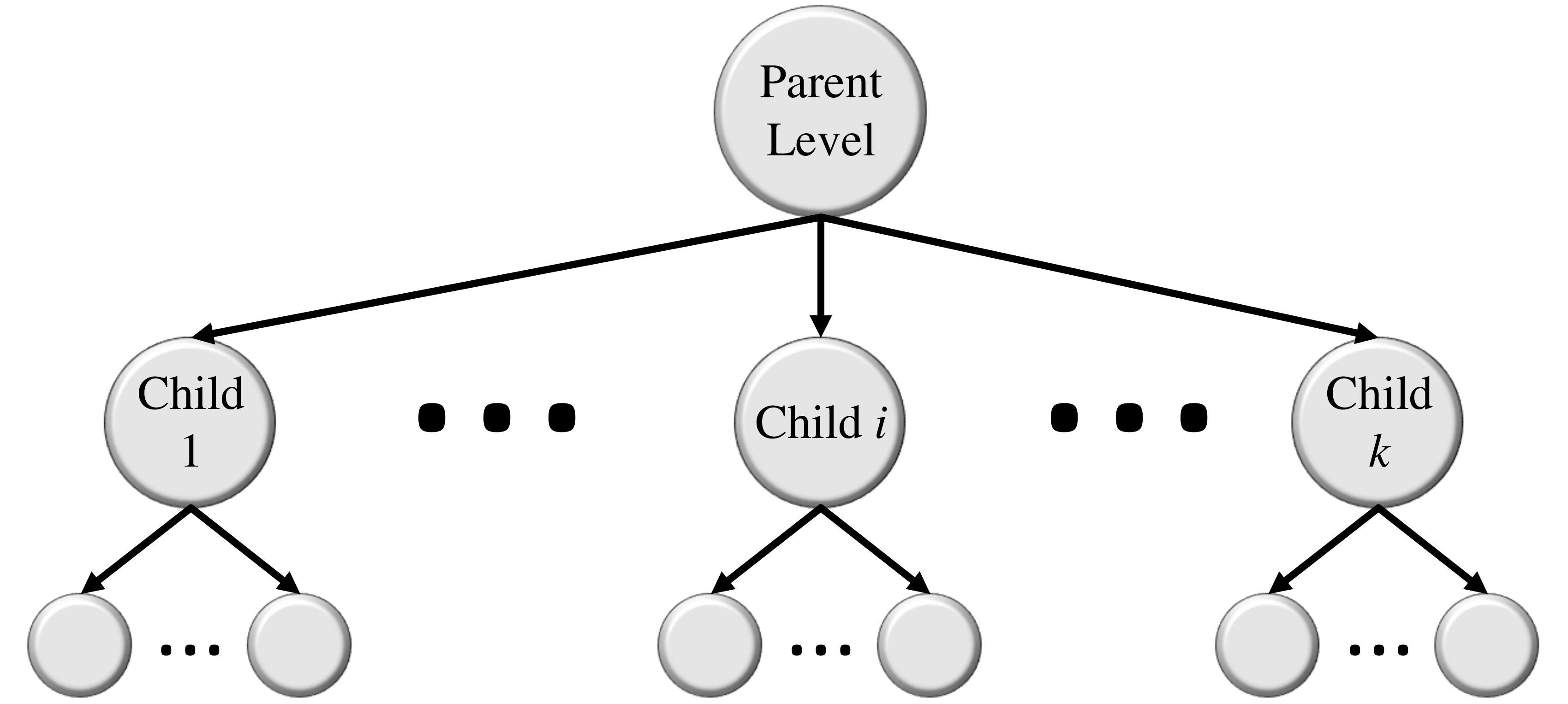}
    \caption{HMIC: Hierarchical Medical Image Classification} \label{fig:HMIC}
\end{figure}

The HMIC approach is shown in Figure \ref{fig:HMIC}. The parent level is a model trained based on the parent level of data; EE, CD or Normal. The child level model is trained for sub-classes of CD based on Modified Marsh Score based on severity; I, IIIa, IIIb, and IIIc).

The rest of this paper is organized as follows: In Section~\ref{sec:Data_Source}, the different data sets used in this work, as well as, the required pre-processing steps are described. The architecture of the model is explained in Section~\ref{sec:Method}. Empirical results are elaborated in Section~\ref{sec:Empirical_Results}. Finally, Section~\ref{sec:Conclusion} concludes the paper along with outlining future directions.

\section{Data Source}\label{sec:Data_Source}As shown in Table~\ref{ta:population}, 
the biopsies were already  obtained from 150 children in this study with a median (interquartile range) age of 37.5 (19.0~to~121.5) months and a roughly equal sex distribution; $77$~males~$(51.3\%)$, and LAZ/ HAZ (Length/ Height-for-Age Z score) of the EE participants were $-2.8$ (inter-quartile range $(IQR): -3.6$~to~$-2.3)$ and $-3.1$ (IQR:~$-4.1$~to~$-2.2$). LAZ/ HAZ of the Celiac participants were~$-0.3$~(IQR: $-0.8$~to~0.7).  
and LAZ/ HAZ for Normal were~$-0.2$ (IQR: $-1.3$~to~0.5). Duodenal biopsy samples were developed into~$461$ whole-slide biopsy images and labeled as either Normal, EE, or CD. The biopsy slides for EE patients were collected from the Aga Khan University Hospital~(AKUH) in Karachi, Pakistan~($n = 29$ slides from~$10$ patients), and the University of Zambia Medical Center in Lusaka, Zambia ($n = 16$). The slides for Normal patients ($n = 63$) and CD ($n = 34$) were collected from The University of Virginia~(UVa). Normal and CD slides were transformed into a whole-slide at~$40\times$ amplification using the Leica SCN~$400$ slide scanner (Meyer Instruments, Houston, TX, USA) at UVa, and the digitized EE slides of $20\times$ 
and shared by means of the Environmental Enteric Dysfunction Biopsy Investigators~(EEDBI) Consortium shared WUPAX server. The patient populace is as per the following: 

The median age of~($Q1$, $Q3$) of our whole investigation populace was~$37.5$~($19.0$, $121.5$) months, and we had a generally equivalent dispersion of females~($48$\%, $n = 49$) and males~($52$\%, $n = 53$). Most of our examination populace were histologically Normal controls~$(37.7\%)$, followed by CD patients~$(51.8\%)$, and EE patients~$(10.05\%)$.

\begin{table}[H]
\centering
\caption{Population results of biopsies dataset.}\label{ta:population}
\begin{tabular}{ c  c  c  c  c c }
\toprule
                                                                        & \textbf{Total Population      }                                                                          & \textbf{Pakistan                                                                                       } & \textbf{Zambia      }                                                                                    & \multicolumn{2}{c}{\textbf{US}}                                                                                                                                                                   \\ \midrule
Data                                                                    & 150                                                                                             & EE (n = 10)                                                                               & EE (n = 16)                                                                                     & Celiac (n = 63)                                                                             & Normal (n = 61)                                                                             \\ \midrule
\begin{tabular}[c]{@{}c@{}}Biopsy  \\ Images\end{tabular}              & 461                                                                                             & 29                                                                                        & 19                                                                                              & 239                                                                                         & 174                                                                                         \\ \midrule
\begin{tabular}[c]{@{}c@{}}Age,   median \\ (IQR),  months\end{tabular} & \begin{tabular}[c]{@{}c@{}}37.5 \\ (19.0 to 121.5)\end{tabular}                                 & \begin{tabular}[c]{@{}c@{}}22.2 \\ (20.8 to 23.4)\end{tabular}                            & \begin{tabular}[c]{@{}c@{}}16.5 \\ (9.5 to 21.0)\end{tabular}                                   & \begin{tabular}[c]{@{}c@{}}130.0 \\ (85.0 to 176.0)\end{tabular}                            & \begin{tabular}[c]{@{}c@{}}25.0 \\ (16.5 to 41.0)\end{tabular}                              \\ \midrule
\begin{tabular}[c]{@{}c@{}}Gender,  \\ n (\%)\end{tabular}        & \multicolumn{1}{l}{\begin{tabular}[c]{@{}l@{}}M = 77 (\%51.3)\\ 
F  = 73 (\%48.7)\end{tabular}} & \multicolumn{1}{l}{\begin{tabular}[c]{@{}l@{}}M = 5 (\%50)\\ F  = 5 (\%50)\end{tabular}} & \multicolumn{1}{l}{\begin{tabular}[c]{@{}l@{}}M = 10 (\%62.5)\\ F  = 6  (\%37.5)\end{tabular}} & \multicolumn{1}{l}{\begin{tabular}[c]{@{}l@{}}M = 29 (\%46)\\ F  = 34 (\%54)\end{tabular}} & \multicolumn{1}{l}{\begin{tabular}[c]{@{}l@{}}M = 33 (\%54)\\ F  = 28 (\%46)\end{tabular}} \\ \midrule
\begin{tabular}[c]{@{}c@{}}LAZ/  HAZ, \\ median (IQR)\end{tabular}      & \begin{tabular}[c]{@{}c@{}}-0.6 \\ ($-$1.9 to 0.4)\end{tabular}                                   & \begin{tabular}[c]{@{}c@{}}$-$2.8\\  ($-$3.6 to -2.3)\end{tabular}                            & \begin{tabular}[c]{@{}c@{}}$-$3.1 \\ ($-$4.1 to$ -$2.2)\end{tabular}                                  & \begin{tabular}[c]{@{}c@{}}$-$0.3 \\ ($-$0.8 to 0.7)\end{tabular}                               & \begin{tabular}[c]{@{}c@{}}$-0.2$ \\ ($-1.3$ to 0.5)\end{tabular}                               \\ \bottomrule
\end{tabular}
\end{table}

$239$ Hematoxylin and eosin~(H\&E) stained duodenal biopsy samples were collected from the archived biopsies of $63$ CD patients from the University of Virginia~(UVa) in Charlottesville, VA, United States. The sample were converted into whole-slide images at 40$\times$ magnification using the Leica SCN 400 slide scanner~(Meyer Instruments, Houston, TX, USA) at the Biorepository and Tissue Research Facility at UVa. The median age of the UVa patient populace is $130$ months with interquartile ranges of $85.0$ and $176.0$ months for $Q1$ and $Q3$, respectively. UVa images had a generally equivalent circulation of females $(54\%, n=54)$ and male~$(46\% ,n=29)$. The biopsy labels for this research were determined by two clinical experts and approved by a pathologist with considerable authority in gastroenterology. Our dataset is ranged from Marsh I to IIIc with no biopsy declared as Marsh II.

Based on Table~\ref{ta:data}, the biopsy images are patched in to~91,899 total images which contain 32,393 normal patches, 29,308 EE patches, and 30,198 CD patches. In the child level of the medical biopsy patches, CD contains 4 severities of disease~(Type I, IIIa, IIIb, and IIIc) which has 7125 Type I patches, 6842 Type IIIa patches, $8120$ Type IIIb patches, and 8111 Type IIIb patches. The training set for normal and EE contains 22,676 and 20,516 patches, respectively, and for testing 9717 and 8792 patches, respectively. For CD, we have two sets of training and testing where one belongs to the parent model and the other belongs to child level. The parent set contains 21,140 patches for training and 9058 image patches for testing with the common label of CD for all. In the CD child dataset, we have four severity types of this disease (I, IIIa, IIIb, and IIIc). Type I of CD  contains 4988 patches in the training set and 2137 patches in the test set. Type IIIa of CD contains 4790 patches in the training set and 2052 patches in the test set. Type IIIb of CD contains 5684 patches in the training set and 2436 patches in the test set. Finally,  IIIc of CD contains 5678 patches in the training set and 2137 patches in the test set.
\begin{table}[H]
\caption{Dataset used for Hierarchical Medical Image Classification~(HMIC).}\label{ta:data}
\centering
\begin{tabular}{cccccccc}
\midrule
\multicolumn{2}{c}{\textbf{Data}}                           & \multicolumn{2}{c}{\textbf{Train}}      & \multicolumn{2}{c}{\textbf{Test}}      & \multicolumn{2}{c}{\textbf{Total}}      \\ \midrule
\multicolumn{2}{c}{Normal}                         & \multicolumn{2}{c}{22,676}     & \multicolumn{2}{c}{9717}     & \multicolumn{2}{c}{32,393}     \\ \midrule
\multicolumn{2}{c}{Environmental Enteropathy} & \multicolumn{2}{c}{20,516}     & \multicolumn{2}{c}{8792}     & \multicolumn{2}{c}{29,308}     \\ \midrule
\multirow{5}{*}{Celiac Disease}      &          & Parent                  & Child & Parent                 & Child & Parent                  & Child \\ \cmidrule{2-8} 
                                          & I        & \multirow{4}{*}{21,140} & 4988 & \multirow{4}{*}{9058} & 2137 & \multirow{4}{*}{30,198} & 7125 \\ 
                                          & IIIa     &                         & 4790 &                        & 2052 &                         & 6842 \\ 
                                          & IIIb     &                         & 5684 &                        & 2436 &                         & 8120 \\  
                                          & IIIc     &                         & 5678 &                        & 2433 &                         & 8111 \\ \midrule
\end{tabular}
\end{table}

\section{Pre-Processing}\label{sec:Pre-Processing}
In this section, we explain the entirety of the pre-processing steps which includes medical image patching, image clustering to remove useless information, and color balancing to solve the staining problem. The biopsy images are unstructured, can vary in size, and are often very high resolution to even consider processing with deep neural systems. Therefore, it becomes necessary to tile the whole-slide images into smaller image subsets called patches. Many of the patches created after tiling the whole-slide image will not contain useful biopsy tissue data. For example, some patches only contain the white or light-gray background area. In the image clustering section, the process to select useful images is described. Lastly, color balancing is used to address staining problems which is a typical issue in histological image preparation.

\subsection{Image Patching}\label{subsec:Image_Patching}
Although the effectiveness of CNNs in image classification has been shown in various studies in different domains, training on high-resolution Whole Slide Tissue Images (WSI) is not commonly preferred due to a high computational cost. Applying CNNs on WSI can also lead to losing a large amount of discriminative data because of severe down-sampling~\cite{hou2016patch}. Due to cellular level contrasts between Celiac Disease, Environmental Enteropathy, and Normal cases, an image classification model performed on patches can perform at least similarly to a WSI-level classifier~\cite{hou2016patch}.For this study, patches are labeled with the same class as the associated WSI. The CNN models are trained to predict the presence of disease or disease severity at the patch-level.

\subsection{Clustering}\label{subsec:Clustering}
\begin{figure}[H]
    \centering
    \includegraphics[width=\textwidth]{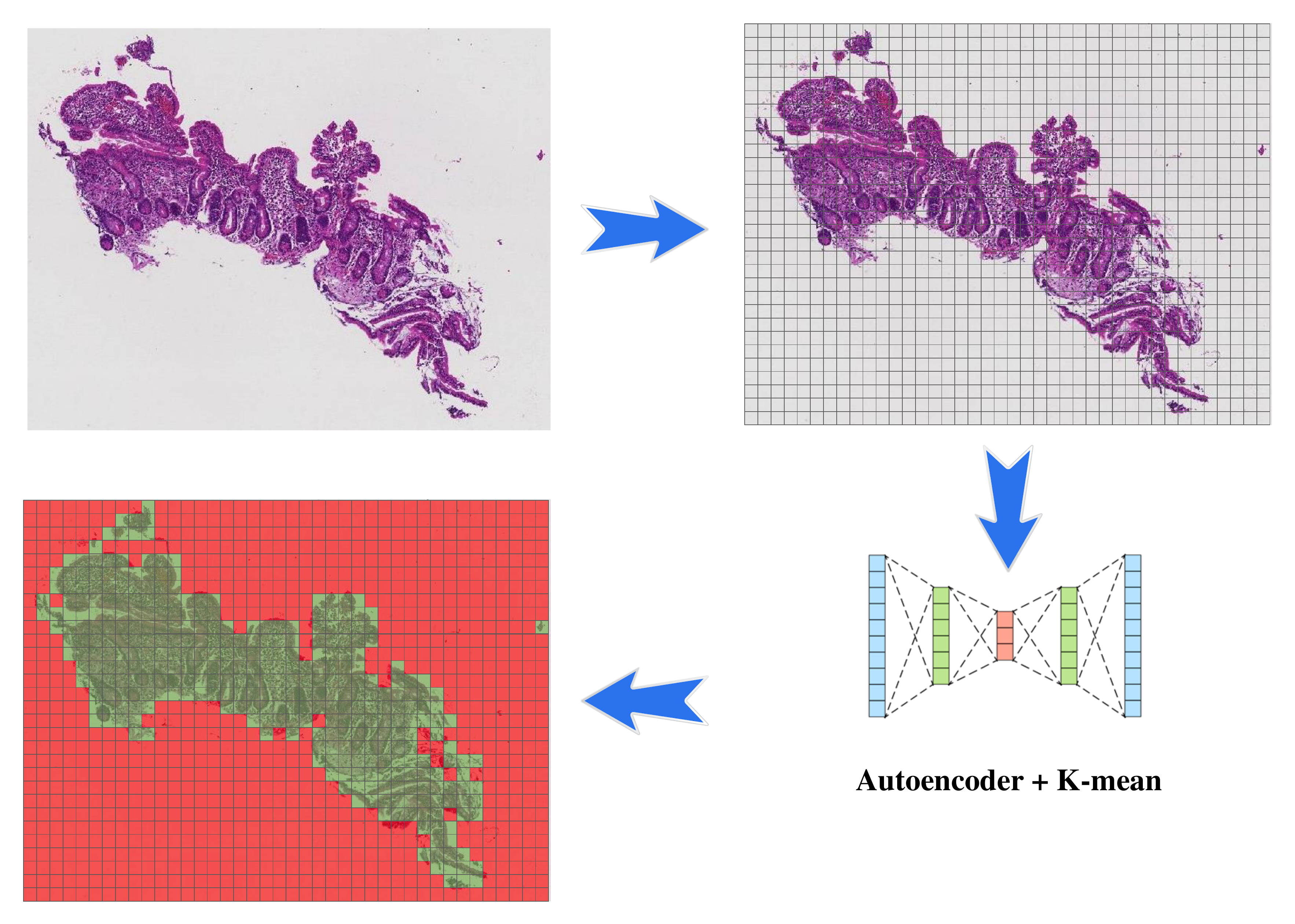}
    \caption{Pipeline of patching and applying an autoencoder to find useful patches for the training model. The biopsy images are very large, so we need to divide into smaller patches to be used in the machine learning model. As you can see in the image, many of these patches are empty. After using an autoencoder, we can apply a clustering algorithm to discard useless patches (green patches contain useful information, while red patches do not).}\label{fig:Patch}
\end{figure}

\begin{figure}[H]
    \centering
    \includegraphics[width=\textwidth]{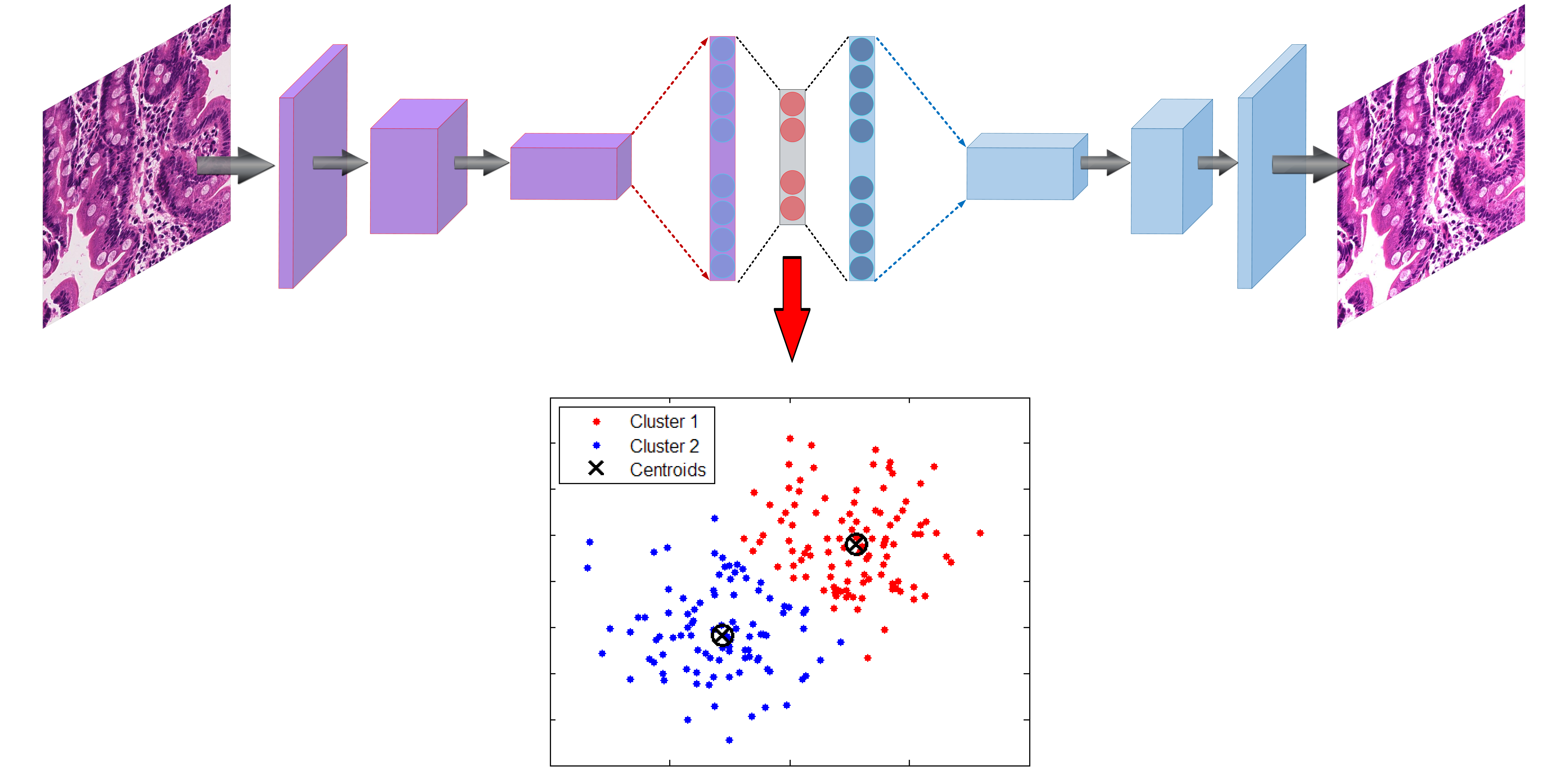}
    \caption{Example autoencoder architecture with K-means applied on the bottle-neck layer feature vector to cluster useful and not useful patches.} \label{fig:AE}
\end{figure}

As shown in Figure~\ref{fig:Patch}, after each biopsy the whole image is divided into patches; many of these patches are not useful input for a deep image classification model. These patches tend to contain only connective tissue, are located on the border region of the tissue, or consist entirely of image background~\cite{kowsari2019diagnosis}. A two-stage clustering process was applied to recognize the immaterial patches. For~the initial step, a convolutional autoencoder was used to learn a vectorized representation of features of each patch and in the second step, we used k-means clustering to assign patches into two groups: helpful and not useful patches. In Figure~\ref{fig:AE}, the pipeline of our clustering strategy is depicted which contains both the autoencoder and \emph{k}-means clustering.

\subsubsection{Autoencoder}

An autoencoder is a form of a neural network that is intended to output a reconstruction of the model's input \cite{goodfellow2016deep}. The autoencoder has achieved incredible success as a dimensionality reduction technique~\cite{wang2014generalized}. The primary version of the autoencoder was presented by~DE. Rumelhart et al.~\cite{rumelhart1985learning} in 1985. The fundamental concept is that one hidden layer acts as a  bottle-neck and has far fewer nodes than other layers in the model~\cite{liang2017text}. This condensed hidden layer can be used to represent the important features of the image with a smaller amount of data. With image inputs, autoencoders can convert the unstructured data into feature vectors that can be processed through other machine learning methods such the k-means clustering algorithm.

\paragraph*{Encode}
A CNN-based autoencoder can be isolated into two principle steps~\cite{masci2011stacked}: encoding and interpreting. This condition is:

\begin{equation}
\begin{split}
    O_m(i, j) = a\bigg(&\sum_{d=1}^{D}\sum_{u=-2k-1}^{2k+1}\sum_{v=-2k -1}^{2k +1}F^{(1)}_{m_d}(u, v)I_d(i -u, j -v)\bigg) \\&\quad m = 1, \cdots, n
\end{split}
\end{equation}
where~$F \in \{F^{(1)}_{1},F^{(1)}_{2},\ldots,F^{(1)}_{n},\}$ is a convolutional filter, with convolution among an input volume defined by~$I = \left\{I_1,\ldots, I_D\right\}$ which it learns to represent the input by combining non-linear functions:
\begin{equation}
    z_m = O_m = a(I * F^{(1)}_{m} + b^{(1)}_m) \quad m = 1, \ldots, m
\end{equation}
where~$b^{(1)}_m$ is the bias, and the number of zeros we want to pad the input with is such that: \text{dim}(I) = \text{dim}(\text{decode}(\text{encode}(I))). Finally, the encoding convolution is equal to:
\begin{equation}
\begin{split}
     O_w = O_h &= (I_w + 2(2k +1) -2) - (2k + 1) + 1 \\&= I_w + (2k + 1) - 1
\end{split}
\end{equation}

\paragraph{Decode}
The decoding convolution step produces~$n$ feature maps~$z_{m=1,\ldots,n}$. The reconstructed results~$\hat{I}$ is the result of the convolution between the volume of feature maps~$Z=\{z_{i=1}\}^n$ and this convolutional filters volume~$F^{(2)}$~\cite{chen2015page,geng2015high}.
\begin{equation}
    \tilde{I} = a(Z * F^{(2)}_{m} + b^{(2)})
\end{equation}
\begin{equation}\label{eq:a:CNN}
\begin{split}
      O_w = O_h &= ( I_w + (2k + 1) - 1 ) -  (2k + 1) + 1 = I_w =I_h
\end{split}
\end{equation}
where Equation~\eqref{eq:a:CNN} shows the decoding convolution with ~$I$ dimensions. The input's dimensions are equal to the output's dimensions.

\subsubsection{K-Means}
\label{sec:kmeans}
K-means clustering is one of the most popular clustering algorithms~\cite{jain2010data,gao2019soft,kowsari2015construction,kowsari2016weighted,alassaf2015automatic} for data in the form $D\in\{x_1,x_2,\ldots,x_n\}$ in $d$ dimensional vectors for $x\in f^d$. K-means had been applied to perform image and data clustering for information retrieval~\cite{jain2010data,manning20introduction,10.1007/978-3-642-00202-1_24}. The aim is to identify groups of similar data points and assign each point to one of the groups. There are many other clustering algorithms, but the k-means approach works well for this problem, because there are only two clusters and it is computationally inexpensive compared to other methods.

As an unsupervised approach, one measure of effective clustering is to sum the distances of each data point from the centroids of the assigned clusters. The goal of K-means is to minimize $\xi$, the sum of these distances, by determining optimal centroid locations and cluster assignments.  This algorithm can be difficult to optimize due to the volatility of cluster assignments as the centroid locations change. Therefore, the K-means algorithm is a greedy-like approach that iteratively adjusts these locations to solve the minimization.

Minimize~$\xi$ with respect to~$A$ and~$\mu$ by:
\begin{align}
    \xi = \sum_{j=1}^k \sum_{x_i} ||x_i-\mu_j||^2 = \sum_{j=1}^k \sum_{i=1}^n A_{ij}||x_i-\mu_j||
\end{align}
where $x_i$ are values from the autoencoder feature representation, $\mu_j$ is the centroid of each cluster, and $A_{ij}$ is the cluster assignment of each data point $i$ with cluster $j$. $A_{ij}$ can only take on binary values and each data point can only be assigned to a single cluster.

The centroid $\mu$ of each cluster is calculated as follows:
\begin{equation}
    \mu (w) = \frac{1}{|w|} \sum_{\bar{x}\in w} \bar{x}
\end{equation}

\begin{figure}[H]
    \centering
    \includegraphics[width=0.7\columnwidth]{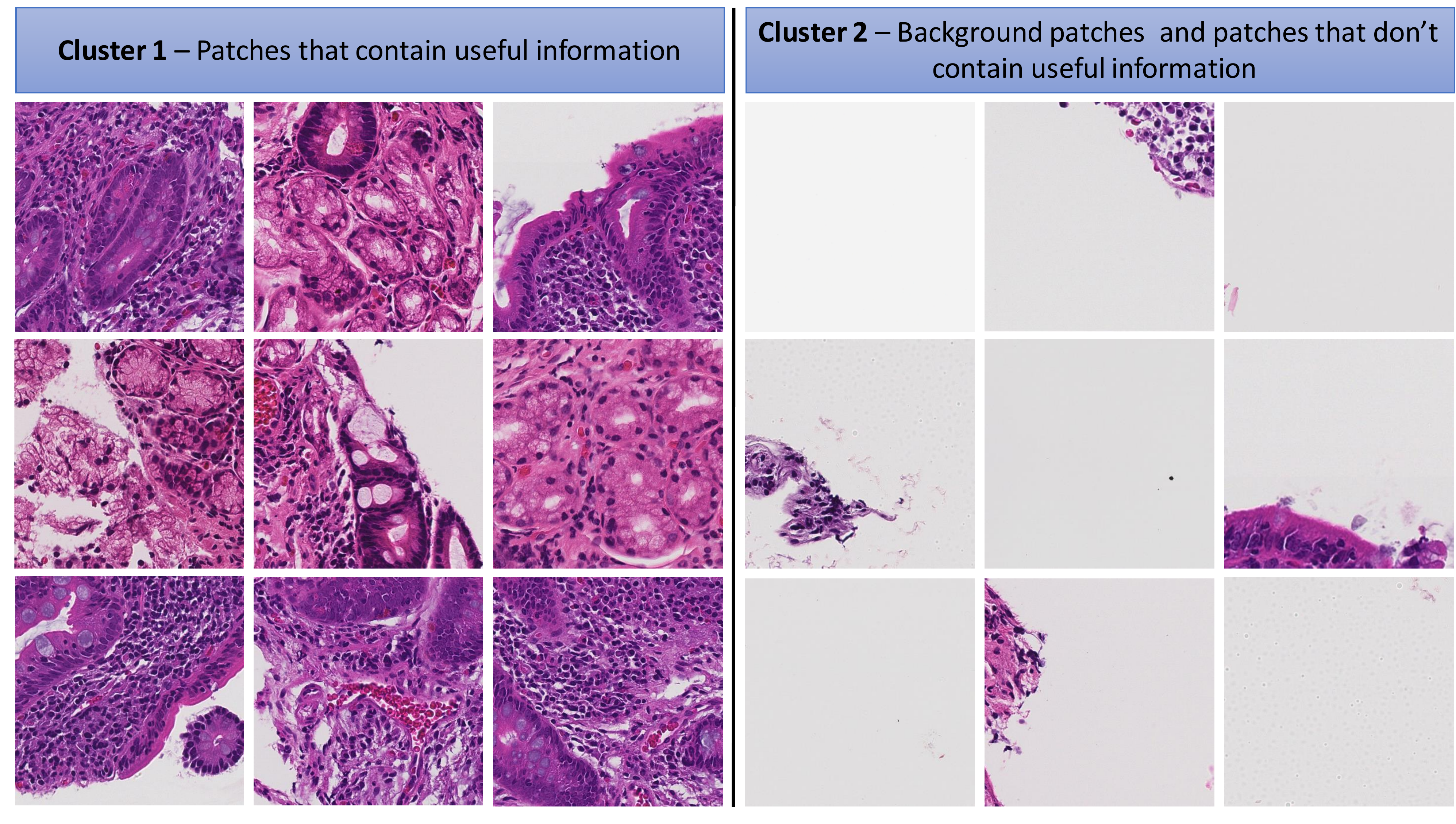}
    \caption{Some samples of clustering results---cluster 1 includes patches with useful information and cluster 2 includes patches   without useful information (mostly created from background parts of WSIs)} \label{fig_Clustering}
\end{figure}

Finally, as shown in Figure~\ref{fig_Clustering}, all patches are assigned into two clusters which one of them contains useful information and the other one is empty or does not have medical information. The Algorithms~\ref{al.kmeans} indicates kmeans algorithm for two clusters medical images.

\begin{algorithm}[H]
\SetKwFor{Foreach}{for}{do}{endfor}
\SetKwFor{WHILE}{while}{do}{endwhile}
\caption{{K-means algorithm for 2 clusters medical images}}\label{al.kmeans}

\textbf{Input: }$D= \{\overrightarrow{x_1},\overrightarrow{x_2},\ldots,\overrightarrow{x_n}\}$

\textbf{Output: } $\mu = \{ \overrightarrow{\mu_1},\overrightarrow{\mu_2}\} $

$S = \{\overrightarrow{s_1}, \overrightarrow{s_2}\}$ set random seeds\\ $(\{\overrightarrow{x_1},\overrightarrow{x_2},\ldots,\overrightarrow{x_n}\},K)$

    \Foreach {$i \leftarrow 1$ to K}{
    
$\overrightarrow{\mu_i} \leftarrow \overrightarrow{s_i}$
    
    }
    \WHILE{Criterion has not been met}
    {
      \Foreach {$i \leftarrow 1$ to K=2}{
      
      $w_k \leftarrow \{\}$
      }
      
    \Foreach {$n \leftarrow 1$ to $N$}{
      
      $j \leftarrow arg \min_{j'} |\overrightarrow{\mu_{j'}} - \overrightarrow{\mu_{x_n}}|$

      $w_j \leftarrow w_j \bigcup ~\{\overrightarrow{x_n}\}$

     }
    \Foreach {$i \leftarrow 1$ to K=2}{
      
     $\mu_i \leftarrow \frac{1}{|w_i|} \sum_{\overrightarrow{x}\in w_i} \overrightarrow{x}$
      
      }
    }
\end{algorithm}

\subsection{Medical Image Staining}

Hematoxylin and eosin (H\&E) stains have been used for at least a century and are still essential for recognizing various tissue types and the morphologic changes that form the basis of contemporary CD, EE, and cancer diagnosis~\cite{fischer2008hematoxylin}.  H\&E is used routinely in histopathology laboratories as it provides the pathologist/researcher a very detailed view of the tissue \cite{anderson2011introduction}.
Color variation has been a very important problem in histopathology based on light microscopy. A range of factors makes this problem even more complex such as the use of different scanners, variable chemical coloring/reactivity from different manufacturers/batches of stains, coloring being dependent on staining procedure (timing, concentrations, etc.), and light transmission being a function of section thickness~\cite{khan2014nonlinear}. Different H\&E staining appearances within machine learning inputs can cause the model to focus only on the broad color variations during training. For example, if images with a certain label all have a unique stain color appearance, because they all originated from the same location, the machine learning model will likely leverage the stain appearance to classify the images rather than the important medical cellular~features.

\subsubsection{Color Balancing}\label{subsec:CB}
The idea of color balancing for this study is to convert images in to a similar color space to represent variations in H\&E staining. The images can be represented with the illuminant spectral power distribution as shown by~$I(\lambda)$, the surface spectral reflectance~$S(\lambda)$, and the~$C(\lambda)$ is sensor spectral sensitivities~\cite{bianco2017improving,bianco2014error}. Using these notations~\cite{bianco2014error}, the sensor reactions at the pixel with coordinates of~$(x,y)$  which can be presented as:

\begin{equation}
    p(x,y) = \int_w I(x,y,\lambda) S(x,y,\lambda) C(\lambda) d\lambda
\end{equation}
where~$w$ is the wavelength range of the visible light spectrum, p and~$C(\lambda)$ are three-component vectors.

\begin{equation}\label{eq_RGB_IN_OUT}
    \begin{aligned} \left [ \begin{array}{c} R \\ G \\ B \\ \end{array} \right ]_{out} =& \left( \alpha \left [ \begin{array}{c@{\quad}c@{\quad}c} a_{11} & a_{12} & a_{13} \\ a_{21} & a_{22} & a_{23} \\ a_{31} & a_{32} & a_{33}\\ \end{array} \right ]\right. {}\times\left. \left [ \begin{array}{c@{\quad}c@{\quad}c} r_{i} & 0 & 0 \\ 0 & g_{i} & 0 \\ 0 & 0 & b_{i} \\ \end{array} \right ] \left [ \begin{array}{c} R \\ G \\ B \\ \end{array} \right ]_{in} \right)^{\gamma} \end{aligned}
\end{equation}
where\textbf{~$RGB_{in}$} stand for the raw images from medical images, and  the diagonal matrix diag($r_{i},g_{i},b_{i}$) is the channel-independent gain compensation of the illuminant~\cite{bianco2014error}. In addition,~\textbf{~$RGB_{out}$} is output results that be send to input feature space of CNN models. $\gamma$ is the gamma correction defined for the RGB color space
and RGB$_{out}$ are the output RGB values. In the following, a more compact version of Equation~\eqref{eq_RGB_IN_OUT} is used:
\begin{equation}
    RGB_{out} = (\alpha AI_w . RGB_{in})^\gamma
\end{equation}

where~$\alpha$ stand for exposure compensation gain, and the diagonal matrix for the illuminant compensation shows by~$I_w$ and the color matrix transformation is shown by matrix $A$ which is a diagonal matrix for the illuminant compensation and the color matrix transformation~\cite{bianco2014error}. 

\begin{figure}[H]
    \centering
    \includegraphics[width=\columnwidth]{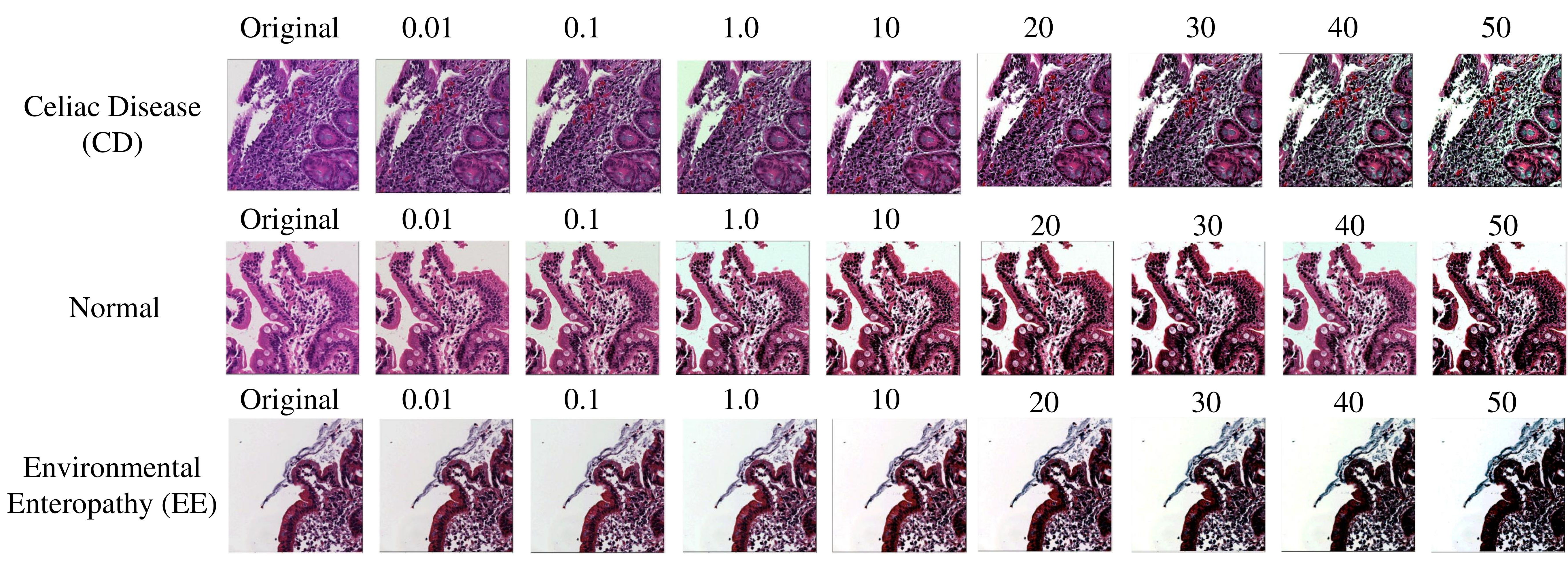}
    \caption{Color Balancing samples for the three classes.} \label{fig:CB}
\end{figure}

Figure~\ref{fig:CB} indicates the output results of three classes~(CD, EE, and Normal) for color balancing (CB)  with various color balancing percentage in range between~$0.01$ and~$50$.

\subsubsection{Stain Normalization}\label{subsec:StainNormalization}

\begin{figure}[H]
    \centering
    \includegraphics[width=\columnwidth]{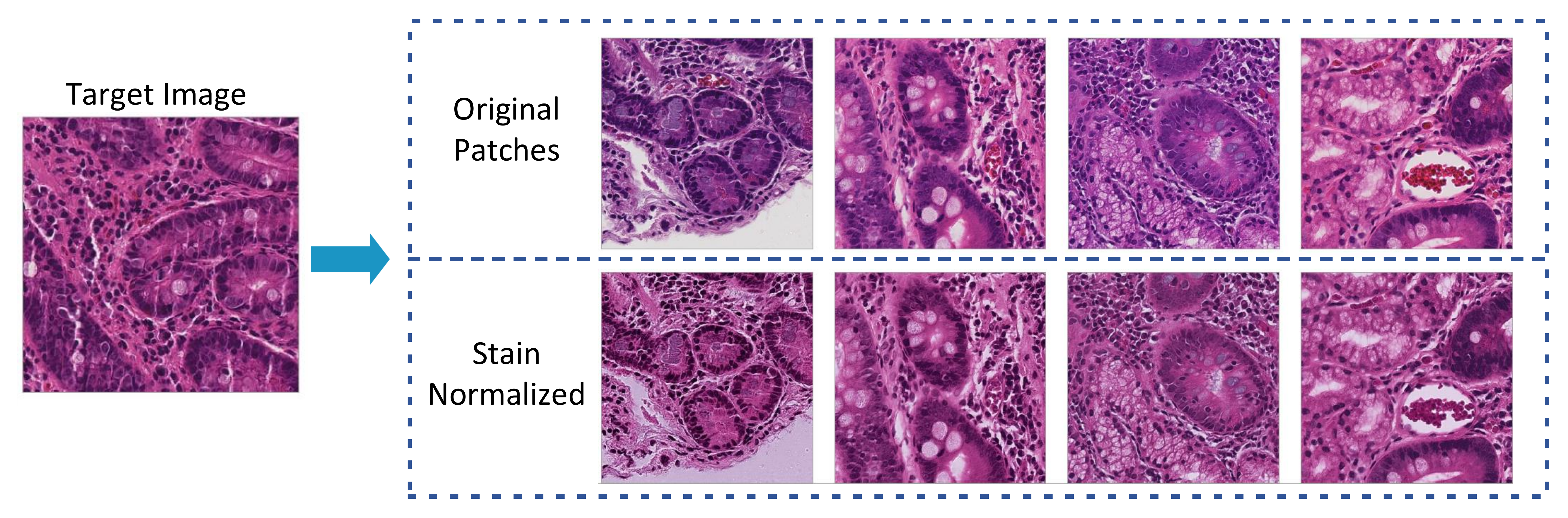}
    \caption{Stain normalization results when using the method proposed by Vahadane et al.~\cite{vahadane2016structure}. Images in the first row represent the source images. The source images are normalized images to the stain appearance of the target image in second row~\cite{sali2019celiacnet}.} \label{fig:StainNormalization}
\end{figure}

Histological images can have significant  variations in stain appearance that will cause biases during model training~\cite{sali2019celiacnet}. The variations occur due to many factors such as contrasts in crude materials and assembling procedures of stain vendors, staining conventions of labs, and color reactions to digital scanners~\cite{vahadane2016structure,sali2019celiacnet}. To solve this problem, the stains of all images are normalized to a single stain appearance. Different staining normalization approaches have been proposed in research projects. In this paper, we used the methodology proposed by Vahadane et al.~\cite{vahadane2016structure} for the CD severity child-level since all images are collected from one center. This methodology is designed to preserve the structure of cellular features of images after stain normalization and accomplishes stain separation with non-negative matrix factorization. Figure~\ref{fig:StainNormalization} shows an example outputs before and after applying this method on biopsy patches.

\section{Baseline}
\unskip
\subsection{Deep Convolutional Neural Networks}
A Convolutional Neural Network~(CNN) performs hierarchical medical image classification for each individual image. The original version of the CNN was built for image processing with an architecture similar to the visual cortex. In this basic CNN baseline for image processing, an~image tensor is convolved with a set of $d \times d$ kernels size. These convolution layers are called feature maps and these provide multiple filters which could be stacked on the input. We used a flat CNN~(non-hierarchical CNN) as one of our baselines.

\subsection{Deep Neural Networks}
A Deep Neural Network (DNN) or multilayer perceptron is designed to be trained by multiple layers of connections. Each individual hidden layer can receive connection from the previous hidden layers' nodes and only can provide connections to the next layer. The input is a connection of flattened feature space~(RGB). The output layer is number of classes for multi-class classification~(six nodes). Our baseline implementation of DNN~(multilayer perceptron) is a discriminative trained model that uses a standard back-propagation algorithm with sigmoid ~(Equation~\eqref{sigmoid}) and Rectified Linear Units (ReLU)~\cite{nair2010rectified}~(Equation~\eqref{relu}) activation functions. The output layer for classification task uses the $Softmax$ function due to having multi-class output as shown in Equation~\eqref{Softmax}.

\section{Method}\label{sec:Method}

\begin{figure}[H]
    \centering
    \includegraphics[width=0.91\textwidth]{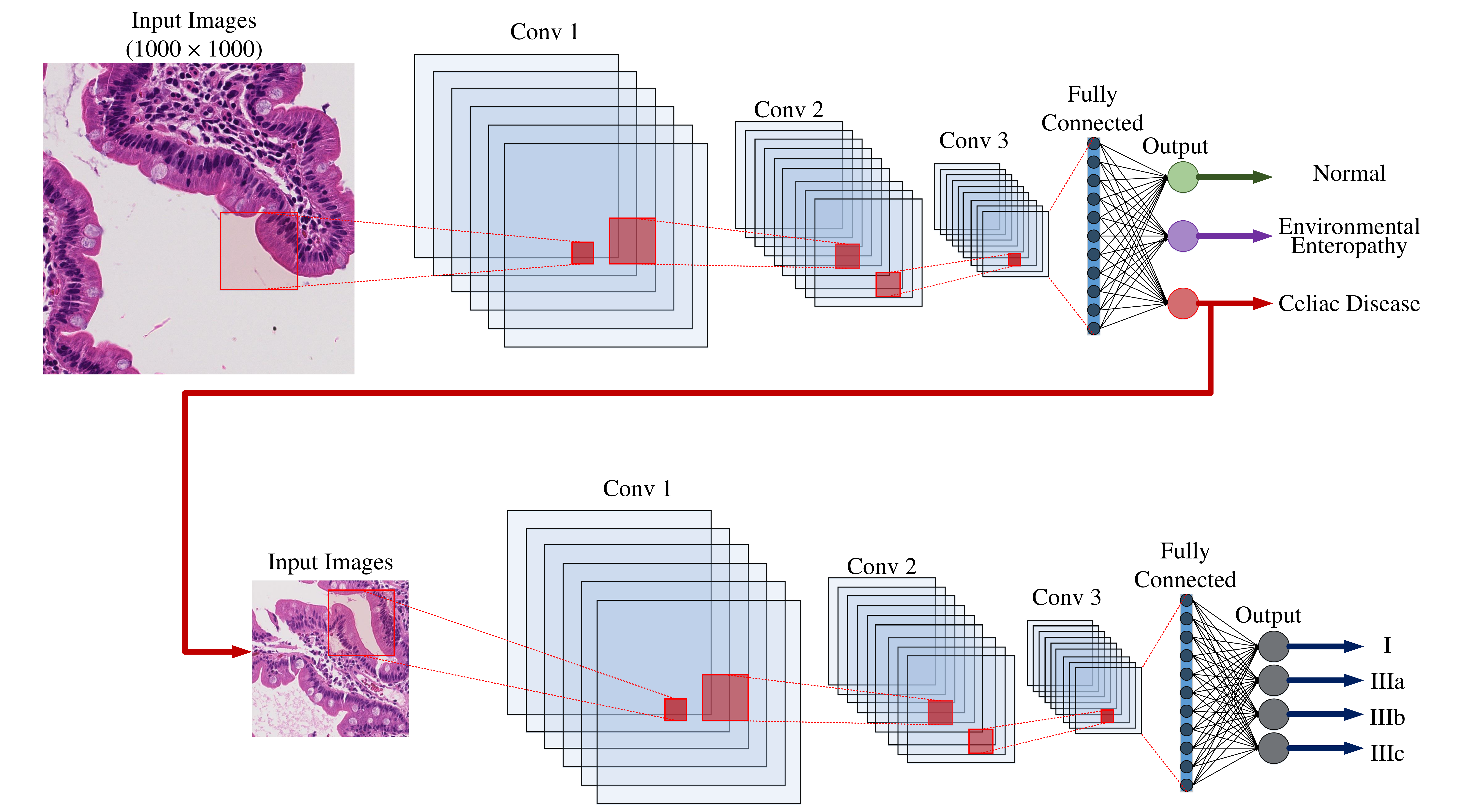}
    \caption{Structure of Convolutional Neural Net using multiple 2D feature detectors and 2D max-pooling} \label{fig:cnn}
\end{figure}

In this section, we explain our concept of Deep Convolutional Neural Networks~(CNN) containing the convolutional layers, activation functions, pooling-layers, and finally, the optimizer. Then, we describe our Deep Convolutional Neural Networks architecture to diagnose Celiac disease and environmental enteropathy. As shown in Figure~\ref{fig:cnn}, the input layer consists of image patches with size of~($1000 \times 1000$ pixels) and it follows the connection to the convolutional layer~(\textit{Conv~$1$}). Conv~$1$ connects to the its following pooling layer~(\textit{MaxPooling}). The pooling layer is connected to second convolutional layer \textit{Conv~$2$}. The last convolutional layer~(\textit{Conv~$3$}) has been flattened and connected to a fully connected multi-layer perceptron. The final layer includes three nodes where each individual node represents one class.
\subsection{Convolutional Neural Networks}\unskip
\subsubsection{Convolutional Layer}
Convolutional Neural Networks are deep learning models that can be used for the hierarchical classification tasks, especially, image classification~\cite{kowsari2018rmdl}. Initially, CNNs were designed for image and computer vision with a  similar design as the visual cortex. CNNs have been used successfully for clinical image classification. In CNNs, an image tensor is convolved with set of~$d \times d$ kernels. These convolutions~(``Feature Maps'') can be stacked to represent many different features detected by the filters in that layer. The feature dimensions of output and input networks can be different~\cite{li2014medical}. The~procedure for processing a solitary output of a matrix is characterized as follows:
\begin{equation}
    A_{j}=f\left(\sum_{i=1}^{N}I_{i}\ast K_{i,j}+B_{j}\right)
\end{equation}
Each individual matrix~$I_i$ is convolved with its corresponding kernel matrix~$K_{i,j}$, and bias of~$B_j$. Finally, a activation function~(non-linear activation function is explained in Section~\ref{Sec:Activation}) is applied to each individual element~\cite{li2014medical}.

The biases and weights are adjusted to constitute competent feature detection filters after the back-propagation step during CNN training. The feature map filters are applied across all three channels~\cite{Heidarysafa2018RMDL}.

\subsubsection{Pooling Layer}
To diminish the computational multifaceted nature, CNNs use pooling layers which decrease the size of the output layer from its input with one layer then onto the next in the networks. Distinctive pooling procedures are used to decrease output while safeguarding significant features \cite{scherer2010evaluation}. The most widely recognized pooling technique is a max-pooling technique where the largest activation is chosen in the pooling window. 

\subsubsection{Neuron Activation}\label{Sec:Activation}
The CNN is implemented as a discriminative method that uses a back-propagation algorithm derived from sigmoid (Equation \ref{sigmoid}), or (Rectified Linear Units (ReLU) \cite{nair2010rectified} (Equation \ref{relu}) activation functions. The final layer contains one node with sigmoid activation function for binary classification multiple nodes for each class and a $Softmax$ activation function for multi-class problems~(as demonstrated in Equation~\eqref{Softmax}).

 \begin{align}
f(x) &= \frac{1}{1+e^{-x}}\in (0,1)\label{sigmoid}\\
f(x) &= \max(0,x)\label{relu}\\
\sigma(z)_j &= \frac{e^{z_j}}{\sum_{k=1}^K e^{z_k}}\label{Softmax}\\ 
&\forall   ~j \in \{1,\hdots, K\} \nonumber
\end{align}

\subsubsection{Optimizer}\label{sec:optimizer}
For our CNN architecture, we use the $Adam$ optimizer~\cite{kingma2014adam}. This is a stochastic gradient descent that uses the norm of the initial two moments of gradient~($v$ and $m$, appeared in Equations~\eqref{adam}--\eqref{adam3}). It can deal with non-stationarity of the target in a similar fashion to RMSProp, while defeating the sparse gradient problem constraint of RMSProp~\cite{kingma2014adam}.

\begin{equation}
\theta  \leftarrow \theta - \frac{\alpha}{\sqrt{\hat{v}}+\epsilon} \hat{m}\label{adam}
\end{equation}
\begin{equation}
g_{i,t} =  \nabla_\theta J(\theta_i , x_i,y_i) \label{adam1}
\end{equation}
\begin{equation}
m_t = \beta_1 m_{t-1} + (1-\beta_1)g_{i,t}\label{adam2}
\end{equation}
\begin{equation}
m_t = \beta_2 v_{t-1} + (1-\beta_2)g_{i,t}^2\label{adam3}
\end{equation}
where $m_t$ is the first moment and $v_t$ indicates second moment that both are estimated. $\hat{m_t}=\frac{m_t}{1-\beta_1^t}$ and $\hat{v_t}=\frac{v_t}{1-\beta_2^t}$.

\subsubsection{Network Architecture}

As demonstrated in Figure~\ref{fig:cnn}, our implementation contains three convolutional layers with each followed by a pooling layer~(Max-Pooling). This method with three channel input image patches with size a of~$(1000\times 1000$~pixels). The first convolutional layer has~$32$ filters with kernel size of~$(3, 3)$. Then, a pooling layer is connected with size of~$(5,5)$ to reduce feature maps from~$(1000\times 1000)$ to~$(200 \times 200)$. The next convolutional layer includes~$32$ filters with $(3, 3)$~kernel. Then, a~$2D$ MaxPooling layer is connected to scales down the feature space from~$(200\times 200)$ to~$(40 \times 40)$. The final convolutional layers contain~$64$ filters that kernel size is~$(3, 3)$. This convolutional layer is connected to a~$2D$ MaxPooling to scale down by~$(8 \times 8)$. The feature map is flattened, and a fully connected layers is connected to our CNN with~$128$ nodes. The output layer has~$3$ nodes that represent our parent classes:~(Environmental Enteropathy, Celiac Disease, and Normal). The child level of this model as shown on the bottom of Figure~\ref{fig:cnn}, is similar to parent level with significant difference which is that the output layer has~$4$ nodes that represent our child classes:~(I, IIIa, IIIb, and IIIc).

The Adam~(See Section~\ref{sec:optimizer}) optimizer is used with a learning rate of~$0.001$, $\beta_1=0.9$, and $\beta_2=0.999$.  The loss function is sparse categorical crossentropy~\cite{chollet2015keras}. Also, for all layers, we use a Rectified linear unit~(ReLU) as the activation function except for the output layer which used a~$Softmax$~(See Section~\ref{Sec:Activation}). In this technique, we use dropout in each individual layer to address over-fitting problem~\cite{srivastava2014dropout}

\subsection{Whole Slide Classification}\label{sec:WSIClassification}
The objective of this study was to group WSIs dependent on the diagnosis of CD and EE, and CD severity on child-level by means of the adjusted Marsh score. The model was used by training it on the patch-level and is extended to WSI. To accomplish this objective, a heuristic strategy was created which aggregated crop classifications and translated them to whole-slide inferences. Each WSI in the test set was at firstly patched, those patches which did not contain any useful information were filtered out, and then stain methods were performed on the patches~(color balancing applied on parent level and stain normalization applied for CD severity ). After these pre-processing steps, our prepared model was applied with the objective of image classification. We meant the likelihood dissemination over potential marks, given the patches images~$x$ and training set~$D$ by~$p(y\vert x,D)$. Finally, this classification produces a vector of length $C$, where $C$ is the number of classes. In our documentation, the likelihood is contingent on the test patch $x$, just as, the training set~$D$. The trained model predicts a vector of probabilities (three for parent-level and four for child-level) that represents the likelihood an image belongs in each class. Given a probabilistic result, the patch~$j$ in slide $i$ is assigned to the most likely class label~$\hat{y}_{ij}$ as shown in Equation~\eqref{eq:patchClass}.

\begin{equation}\label{eq:patchClass}
\hat{y}_{ij} = \argmax_{c \in \{1,2,3,\ldots,C\}} p(y_{ij} = c\vert x_{ij},D)
\end{equation}

where $\hat{y}$ stands for maximum a posteriori~(MAP). The summation over these vectors~(output vector of all patches for a single WSI) and normalizing the resultant vector made a vector that had parts demonstrating the likelihood of a vector with three elements~(CD, EE, and N) seriousness for the related WSI. Equation~\eqref{eq:slideClass}, shows how the class of WSI was anticipated.

\begin{equation}\label{eq:slideClass}
\hat{y}_i = \argmax_{c \in \{1,2,3,\ldots,C\}} \sum_{j=1}^{N_i}p(y_{ij} = c\vert x_{ij},D)
\end{equation}

where the number of patches in slide $i$ is shown by $N_i$.

\subsection{Hierarchical Medical Image Classification}
The main contribution of this paper is a hierarchical medical image classification of biopsies. A~common multi-class algorithm is functional and efficient for a limited number of categories. However, performance drops when we have an unequal number of data-points in our classes. In our deep learning models with various levels, this issue has been solved by creating a hierarchical structure that makes deep learning approaches for their levels of the clinical hierarchy~(e.g., see Figure~\ref{fig:cnn}).

\section{Results}\label{sec:Empirical_Results}
In this section, we have two main results:  empirical results and visualizations for patches. The~empirical results are mostly used for comparing our accuracy with our baseline.
\subsection{Evaluation Setup}\label{sec:Evaluation}
In the computer science community, shareable and commensurate performance measures to assess an algorithm are desirable. However, in real projects, such measures may only exist for a few methods. The extensive problem when assessing the medical image categorization model is the absence of standard data collection agreement. Even if a commonplace method existed, simply choosing disparate training and test sets can introduce divergencies in model achievement~\cite{yang1999evaluation}. Performance measures widely evaluate specific aspects of image classification. In this section, we explain different performance measures and metrics that are used in this research paper. These metrics have been calculated from a~``confusion matrix'' that comprises false negatives~(FN) true negatives~(TN), true positives~(TP), and false positives~(FP)~\cite{lever2016points}. The importance of these four measures may shift depending on the application. The fraction of all correctly predicted over all number of test set samples is the overall accuracy~(Equation~\eqref{eq:acc}). The fraction of correctly predicted over all positives is called precision,~i.e., positive predictive value (Equation~\eqref{eq:pres}). 

\begin{equation}
    accuracy=\frac{(TP+TN)}{(TP+FP+FN+TN)}\label{eq:acc}
\end{equation}
\begin{equation}
Precision = \frac{\sum_{l=1}^LTP_l}{\sum_{l=1}^LTP_l+FP_l}\label{eq:pres}
\end{equation}
\begin{equation}
Recall= \frac{\sum_{l=1}^LTP_l}{\sum_{l=1}^LTP_l+FN_l}\label{eq:recall}
\end{equation}
\begin{equation}
F1 Score =  \frac{\sum_{l=1}^L2TP_l}{\sum_{l=1}^L2TP_l+FP_l+FN_l}
\end{equation}


\subsection{Experimental Setup}\label{sec:Experimental}
The following results were obtained using a combination of central processing units~(CPUs) and graphical processing units~(GPUs). The processing was done on a $Core~i7-9700F$ with $8$ cores and $128 GB$ memory, and the GPU cards were two $Nvidia~GeForce~RTX~2080 Ti$. We implemented our approaches in Python using the Compute Unified Device Architecture~(CUDA), which is a parallel computing platform and Application Programming Interface~(API) model created by $Nvidia$. We also used Keras and TensorFlow libraries for creating the neural networks~\cite{abadi2016tensorflow,chollet2015keras}. 

\subsection{Empirical Results}
In this sub-section, as we discussed in Section~\ref{sec:Evaluation}, we report precision, recall, and F1-score. 

Table~\ref{ta:parent} shows the results of the parent level model trained for classifying between Normal, Environmental Enteropathy (EE) and Celiac Disease (CD). The precision of normal patches is \mbox{$89.97\pm0.5973$} and recall is $89.35\pm0.6133$. The F1-score of normal is $89.66\pm0.6054$. For EE,  precision is $94.02\pm0.4955$, recall is  $97.30\pm0.3385$, F1-score is $95.63\pm0.4270$. The CD evaluation measure for the parent level is as follows: precision is equal to $91.12\pm0.3208$, recall is equal to $88.71\pm0.3569$, and F1-score is equal to $89.90\pm1.2778$.

\begin{table}[H]
\centering
\caption{Result of parent level classifications for normal, environmental enteropathy, and  Celiac disease.}\label{ta:parent}
\begin{tabular}{ccccc}
\toprule
                               & \textbf{Precision} & \textbf{Recall }& \textbf{F1-Score}  \\ \midrule
Normal                         &   89.97  $\pm$  0.59        &   89.35  $\pm$  0.61     &    89.66  $\pm$  0.60          \\ 
\multicolumn{1}{c }{Environmental Enteropathy } &     94.02  $\pm$  0.49      &   97.30  $\pm$  0.33      &     95.63  $\pm$  0.42       \\ 
Celiac Disease           &      91.12  $\pm$  0.32     &   88.71  $\pm$  0.35     &     89.90  $\pm$  1.27        \\ \bottomrule
\end{tabular}
\end{table}

Table~\ref{ta:results_overall} shows the comparison of our techniques with three different baselines. The baseline results from Convolutional Neural Network~(CNN), Deep Neural Network (Multilayer perceptron), and Deep Convolutional Neural Network (DCNN) are using in this results section. Much research has been done in this domain such as ResNet, but these novel techniques can only handle small images such as $250\times 250$. In this dataset, we create $1000\time 1000$ patches, so we could not compare our work with ResNet, AlexNet, etc. Regarding precision, the highest is HMIC whole-slide with a mean of $88.01$ percent and a confidence interval of $0.3841$ followed by HMIC none whole-slide  $84.13$ percent and confidence interval of $0.3751$. The precision of CNN is $76.76\pm0.4985$, multilayer perceptron is $76.19\pm0.5030$, and DCNN is $82.95\pm0.4439$.  Regarding recall, the highest is HMIC whole-slide with a mean of $93.98$ percent and a confidence interval of $0.2811$ followed by HMIC non whole-slide at $93.56$ percent and confidence interval of $0.29.1$. The recall of CNN is $80.18\pm0.4706$, multilayer perceptron is $79.4\pm0.471$, and DCNN is $87.28\pm0.3933$. The highest F1-score is HMIC whole-slide with a mean of $90.89$ percent and a confidence interval of $0.3804$ followed by HMIC non whole-slide with $88.61$ percent and confidence interval of $0.3751$. The recall of CNN is $78.43\pm0.4855$, multilayer perceptron is $77.76\pm0.4911$, and DCNN is $85.06\pm0.4207$.

\begin{table}[H]
\centering
\caption{Results of HMIC with comparison with our baseline}\label{ta:results_overall}
\begin{tabular}{c c c c c }
\toprule
\multicolumn{2}{c}{\textbf{Model} }                  & \textbf{Precision    }               & \textbf{Recall   }                   & \textbf{F1-Score                                     } \\ \midrule
\multirow{3}{*}{Baseline}      & CNN          & 76.76  $\pm$  0.49          & 80.18  $\pm$  0.47          & 78.43  $\pm$  0.48                   \\ \cmidrule{2-5} 
                               & Multilayer perceptron          &            76.19  $\pm$  0.50                 &             79.40  $\pm$  0.47                &             77.76  $\pm$  0.49                                 \\ \cmidrule{2-5} 
                               &  Deep CNN         & 82.95  $\pm$  0.44          & 87.28  $\pm$  0.39          & 85.06  $\pm$  0.42                  \\ \midrule
\multirow{2}{*}{HMIC} &Non Whole slide & 84.13  $\pm$  0.37 & 93.56  $\pm$  0.29 & 88.61  $\pm$  0.37  \\ \cmidrule{2-5} 
& Whole slide  & \textbf{88.01  $\pm$  0.38} & \textbf{93.98  $\pm$  0.28} & \textbf{90.89  $\pm$  0.38} 
                              \\ \bottomrule 
\end{tabular}
\end{table}

Table~\ref{ta:results_per_class} shows the results by each class. For Normal images, the best classifier is DCNN with $95.14\pm0.42$ recall of $94.91\pm0.43$ F1-score of $95.14\pm0.42$. For EE, HMIC is the best classifier. The~whole-slide images classifier for parent level is more robust in comparison with non -whole slide with precision of $94.08\pm0.49$ Recall of $97.33\pm0.42$ F1-score of $98.68\pm0.42$. Although the results of Normal and EE Images are very similar to flat models such as DCNN, but the results of sub-class of CD contains 4 different stages and the margin is very high. The best flat model (non-hierarchical) is DCNN with mean of F1-score of 73.99 for I, 71.63 for IIIa, 77.74 for IIIb, and 75.71 IIIc.

The Table~\ref{ta:results_per_class} indicates the margin for child level is very high even for the non whole-slide level of this dataset. The best results belong to the whole-slide classifier for parent level with precision with $88.73\pm1.34$ for I, $81.19\pm1.65$ for IIIa, $90.51\pm1.24$ for IIIb, $89.26\pm1.31$ for IIIc. The whole-slide classifier for parent level with recall with $85.07\pm1.51$ for I, $81.19\pm1.65$ for IIIa, $90.48\pm1.27$ for IIIb, $90.18\pm1.26$ for IIIc. The results of whole-slide classifier for parent level for recall is $85.07\pm1.51$ for I, $83.72\pm0.78$ for IIIa, $90.48\pm0.61$ for IIIb, $90.18\pm1.26$ for IIIc. Finally, The F1-score for whole-slide classifier for parent level is equal to $86.86\pm1.43$ for I, $82.44\pm1.51$ for IIIa, $90.49\pm1.16$ for IIIb, $89.72\pm1.28$.

\begin{table}[H]
\centering
\caption{Results per-classed of HMIC with comparison with our baseline.}\label{ta:results_per_class}
\begin{tabular}{c c c c c c c }
\midrule
\multicolumn{3}{c }{\textbf{Model}}                                                             &          &\textbf{ Precision} & \textbf{Recall} & \textbf{F1-Score}  \\ \midrule
    \multirow{26}{*}{Baseline} & \multirow{8}{*}{CNN} & \multicolumn{2}{c}{Normal}                    &    87.83 $\pm$ 0.57       &    90.77 $\pm$ 0.65    &      89.28 $\pm$ 0.61       \\ \cmidrule{3-7}  
                               &                      & \multicolumn{2}{c}{Environmental Enteropathy } &     90.93 $\pm$ 0.61      &    82.48 $\pm$ 0.79    &    86.50 $\pm$ 0.71       \\ \cmidrule{3-7}  
                               &                      & \multirow{5}{*}{Celiac Disease }     & I        &    68.37 $\pm$ 1.98       &   68.62 $\pm$  1.96   &    68.50 $\pm$ 1.96       \\ \cmidrule{4-7} 
                               &                      &                                     & IIIa     &  56.26 $\pm$ 1.01         &    56.26 $\pm$ 2.21    &  59.29 $\pm$ 1.95          \\ \cmidrule{4-7} 
                               &                      &                                     & IIIb     &   65.28 $\pm$ 0.97        &    98.28 $\pm$ 2.01    &   66.64 $\pm$ 1.87           \\ \cmidrule{4-7} 
                               &                      &                                     & IIIc     &  62.66 $\pm$ 1.99         &    66.83 $\pm$ 1.99    &   64.68 $\pm$ 2.02           \\ \cmidrule{2-7}  
                               & \multirow{8}{*}{\begin{tabular}[c]{@{}c@{}}Multilayer \\ perceptron\end{tabular}}    & \multicolumn{2}{c}{Normal}                    &      87.97 $\pm$ 0.76     &    81.87 $\pm$ 0.76    &    84.81 $\pm$ 0.71       \\ \cmidrule{3-7}  
                               &                      & \multicolumn{2}{c}{Environmental Enteropathy } &     87.25 $\pm$ 0.69      &    90.18 $\pm$ 0.62    &     88.69 $\pm$ 0.66       \\ \cmidrule{3-7}  
                               &                      & \multirow{5}{*}{Celiac Disease }     & I        &    57.92 $\pm$ 2.07      &    60.74 $\pm$ 2.07    &    59.30 $\pm$ 2.09       \\ \cmidrule{4-7} 
                               &                      &                                     & IIIa     &     62.58 $\pm$ 2.09      &  62.18 $\pm$ 2.09      &    60.89 $\pm$ 2.11     \\ \cmidrule{4-7} 
                               &                 &                                    & IIIb     &     65.00 $\pm$ 1.89      &    66.09 $\pm$ 1.87    &    65.56 $\pm$ 1.88         \\ \cmidrule{4-7} 
                               &                      &                                     & IIIc     &    67.97 $\pm$ 1.85       &    74.85 $\pm$ 1.72    &   71.24 $\pm$ 1.78       \\ \cmidrule{2-7}  
                               &  \multirow{8}{*}{DCNN}& \multicolumn{2}{c}{Normal}                    &     95.14 $\pm$ 0.42      &    94.91 $\pm$ 0.43    &  95.14 $\pm$ 0.42              \\ \cmidrule{3-7}  
                               &                      & \multicolumn{2}{c}{Environmental Enteropathy } &    92.22 $\pm$ 0.55       &     90.62 $\pm$ 0.60   &    91.52 $\pm$ 0.58         \\ \cmidrule{3-7}  
                               &                      & \multirow{5}{*}{Celiac Disease }     & I        &     75.41 $\pm$ 1.82      &   72.63 $\pm$ 1.89      &    73.99 $\pm$ 1.85          \\ \cmidrule{4-7} 
                               &                      &                                     & IIIa     &      70.81 $\pm$ 1.92     &   72.47 $\pm$ 1.93     &   71.63 $\pm$ 1.79          \\ \cmidrule{4-7} 
                               &                      &                                     & IIIb     &    81.08 $\pm$ 0.81       &    74.67 $\pm$ 1.84    &   77.74 $\pm$ 1.65            \\ \cmidrule{4-7} 
                               &                      &                                     & IIIc     &     75.07 $\pm$ 1.83      &    76.37 $\pm$ 1.81    &   75.71 $\pm$ 1.81          \\ \midrule
\multirow{17}{*}{HMIC} 
& \multirow{8}{*}{Non Whole Slide} & \multicolumn{2}{c}{Normal}                                        & 89.97 $\pm$ 0.59  & 89.35 $\pm$ 0.61 & { 89.66 $\pm$ 0.61} \\ \cmidrule{3-7}  
                                &                               & \multicolumn{2}{c}{Environmental Enteropathy}  & {94.02 $\pm$  0.49} & {97.30 $\pm$ 0.33} & {95.63 $\pm$ 0.33} \\ \cmidrule{3-7}  
                                &                                   & \multirow{5}{*}{{Celiac Disease }} & {I}    & {83.25 $\pm$ 1.58} & {80.91 $\pm$ 1.66} & {82.06 $\pm$ 1.62} \\ \cmidrule{4-7} 
                                &                               &                                          & {IIIa} & {80.34 $\pm$ 1.62} & {80.46 $\pm$ 1.71} & {80.40 $\pm$ 1.57} \\ \cmidrule{4-7} 
                                &                               &                                          & {IIIb} & {85.35 $\pm$ 1.49} & {81.77 $\pm$ 1.67} & {83.52 $\pm$ 1.47} \\ \cmidrule{4-7} 
                                &                               &                                          & {IIIc} & {85.54 $\pm$ 1.49} & {82.71 $\pm$ 1.60} & {84.10 $\pm$ 1.55} \\ \cmidrule{2-7}  
                                
                                & \multirow{8}{*}{{Whole Slide}} & \multicolumn{2}{c}{{Normal}}                     & {90.64 $\pm$ 0.57 } & { 90.06 $\pm$ 0.57} & { 90.35 $\pm$ 0.58} \\ \cmidrule{3-7}  
                                &                               & \multicolumn{2}{c}{{Environmental Enteropathy }} & {94.08 $\pm$  0.49} & {97.33 $\pm$ 0.42} & {98.68 $\pm$ 0.42}  \\ \cmidrule{3-7}  
                                &                               & \multirow{5}{*}{{Celiac Disease }} & {I}    & {88.73 $\pm$ 1.34} & {85.07 $\pm$ 1.51} & {86.86 $\pm$ 1.43} \\ \cmidrule{4-7} 
                                &                               &                                          & {IIIa} & {81.19 $\pm$ 1.65} & {81.19 $\pm$ 1.65} & {82.44 $\pm$ 1.51} \\ \cmidrule{4-7} 
                                &                               &                                          & {IIIb} & {90.51 $\pm$ 1.24} & {90.48 $\pm$ 1.27} & {90.49 $\pm$ 1.16} \\ \cmidrule{4-7} 
                                &                               &                                          & {IIIc} & {89.26 $\pm$ 1.31} & {90.18 $\pm$ 1.26} & {89.72 $\pm$ 1.28}                             \\                              \midrule
\end{tabular}
\end{table}


\subsection{Visualization}

\begin{figure}[H]
    \centering
    \includegraphics[width=\columnwidth]{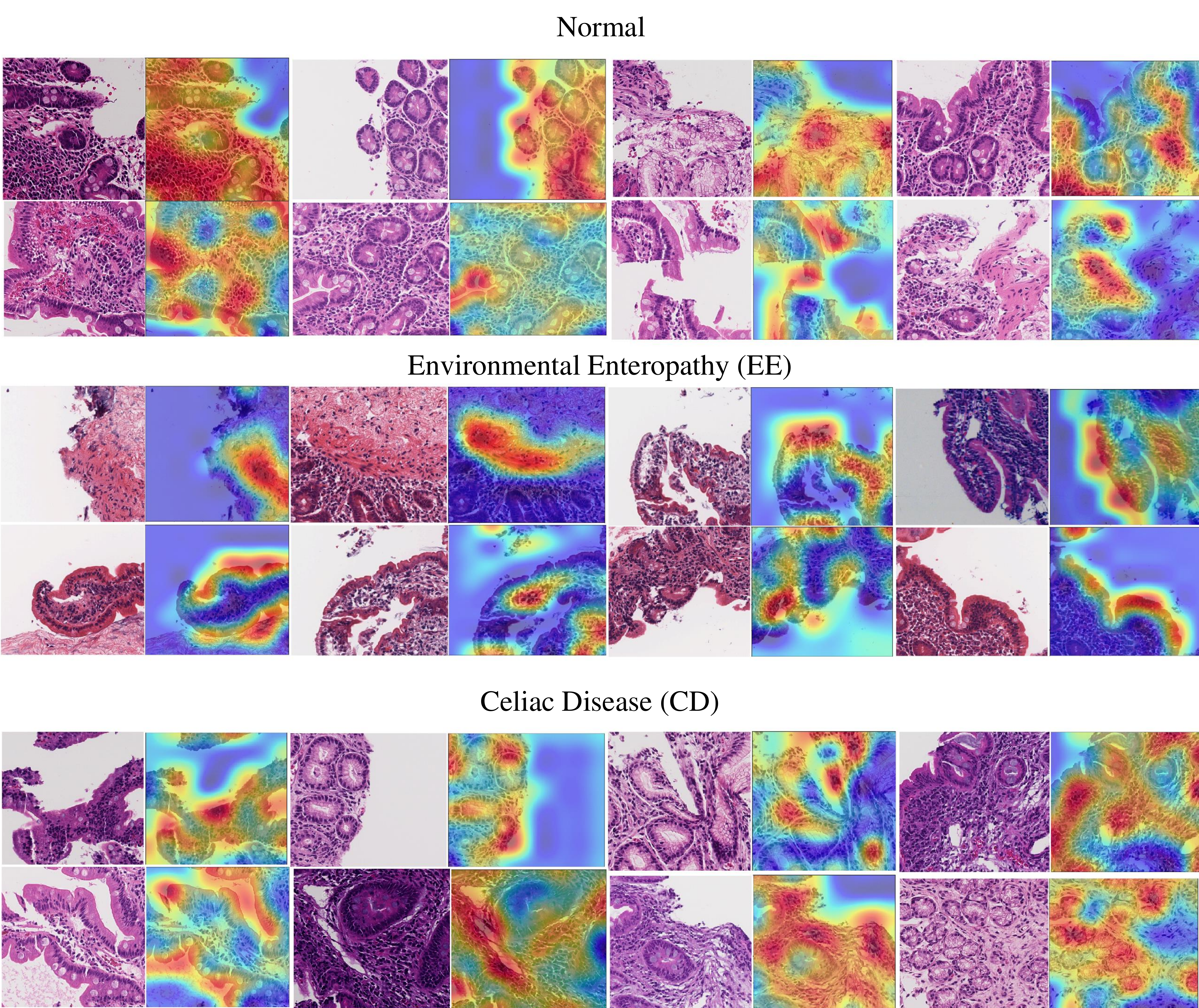}
    \caption{{Grad-CAM} results for showing feature importance.} \label{fig:V_Results}
\end{figure}

Grad-CAMs were generated for 41 patches (18 EE, 14 Celiac Disease, and 9 histologically normal duodenal controls) which mainly focused on distinct, yet medically relevant cellular features outlined below. Although, most heatmaps focused on medically relevant features, there were some patches that focused on too many features (n~=~8) or focused on connective tissue debris (n~=~10) that we were unable to categorize. \\As shown in Figure~\ref{fig:V_Results}, three categories are describe as follows:
\begin{itemize}
    \item EE: surface epithelium with IELs and goblet cells was highlighted. Within the lamina propria, the heatmaps also focused on mononuclear cells. 
	\item CD: heatmaps highlighted the edge of crypt cross sections, surface epithelium with IELs and goblet cells, and areas with mononuclear cells within the lamina propria. 
	\item Histologically Normal: surface epithelium with epithelial cells containing abundant cytoplasm was highlighted. 
\end{itemize}

\section{Conclusions}\label{sec:Conclusion}
Medical image classification is a significant problem to address, given the growing number of medical instruments to collect digital images. When medical images are organized hierarchically, multi-class approaches are difficult to apply using traditional supervised learning methods. This paper introduces a novel approach to hierarchical medical image classification, HMIC, that could use multiple deep convolutional neural networks approaches to produce hierarchical classifications, and in our experimental results, we use two level of CNNs hierarchy. Testing on a medical image data set shows that this technique  produced robust results at the higher and lower level, and the accuracy is consistently higher than those obtainable by conventional approaches using CNN, Multi-layer perceptron, and DCNN. These results show that hierarchical deep learning method could provide improvements for classification and that they provide flexibility to classify these data within a hierarchy. Hence, they provide extensions over current and traditional methods that only consider the multi-class~problem.

This modeling approach can be extended in a couple of ways. Additional training and testing with other hierarchically structured clinical data will help to  identify other architectures that work better for these problems. Also, deeper levels of hierarchy is another possible extension of this approach. For instance, if the stage of the disease is treated as ordered then the hierarchy continues down multiple levels. Scoring here could be performed on small sets using human judges.

\vspace{6pt}

\authorcontributions{K.K., S.S., and DB worked on the Concept and design of the platform. K.K. worked on the implementation of these models. K.K., S.S. , and L.E. worked on the analysis and interpretation of data. K.K. worked on the drafting of the manuscript. K.K., R.S. , and W.A. worked on the critical revision of the manuscript for important intellectual content. D.B , S.S. , B.A. , S.M. and A.A. obtained funding. This work was under the supervision of S.S., P.K., A.A., S.M., and D.B. All authors have read and agreed to the published version of the manuscript.}

\funding{This research was supported by University of Virginia, Engineering in Medicine SEED Grant $(SS~\&~DEB)$, the University of Virginia Translational Health Research Institute of Virginia ($THRIV$) Mentored Career Development Award $(SS)$, and the Bill and Melinda Gates Foundation (AA, OPP1138727; SRM, OPP1144149; PK, OPP1066118. Research reported in this publication was supported by [National Institute of Diabetes and Digestive and Kidney Diseases] of the National Institutes of Health under award number~K23 DK117061-01A1. The content is solely the responsibility of the authors and does not necessarily represent the official views of the National Institutes of Health.}

\conflictsofinterest{The authors declare no conflict of interest. The funding sponsors had no role in the design
of the study; in the collection, analyses or interpretation of data; in the writing of the manuscript; nor in the
decision to publish the results.}

\reftitle{References}


\end{document}